\documentclass[12pt]{article}

\def \a{\alpha}
\def \b{\beta}

\def \d{\delta}
\def \ep{\epsilon}

\def \l{\lambda}
\def \m{\mu}
\def \n{\nu}

\def \s{\sigma}

\def \ph{\phi}

\def \G{\Gamma}
\def \D{\Delta}
\def \T{\Theta}
\def \La{\Lambda}

\def \S{\Sigma}

\def \Om{\Omega}

\def \la#1{\label{#1}}
\def \ift{\infty}
\def \le{\left}
\def \ri{\right}
\def \da{\dagger}
\def \ti#1{\tilde{#1}}
\def \lb{\lbrack}
\def \rb{\rbrack}

\def \rar{\rightarrow}

\def \lrar{\leftrightarrow}
\def \ld{\ldots}
\def \cd{\cdots}
\def \nn{\nonumber}

\newcommand \beq{\begin{eqnarray}}
\newcommand \eeq{\end{eqnarray}}
\newcommand \bea{\begin{eqnarray*}}
\newcommand \eea{\end{eqnarray*}}
\newcommand \ben{\begin{enumerate}}
\newcommand \een{\end{enumerate}}
\newcommand \ba{\begin{array}}
\newcommand \ea{\end{array}}

\begin{document}
\begin{flushright}
MPS-RR 2001-24 \\
\vspace{.4cm}
\end{flushright}

\begin{center}
   {\LARGE\bf A String Bit Hamiltonian Approach} \\
   \vspace{.2cm} 
   {\LARGE\bf to} \\
   \vspace{.3cm}
   {\LARGE\bf Two-Dimensional Quantum Gravity} \\
   \vspace{.7cm}
   {\large {\bf B. Durhuus}$^{a,b,}$\footnote
            {e-mail address: durhuus@math.ku.dk} 
       and {\bf C.-W. H. Lee}$^{a,b,}$\footnote
            {e-mail address: lee@math.ku.dk}} \\
   \vspace{.7cm}
   $^a$ {\it MaPhySto --- Centre of Mathematical Physics and 
   Stochastics\footnote{funded by a grant from the Danish National Research 
   Foundation.}} \\
   $^b$ {\it Department of Mathematics, University of Copenhagen, 
   Universitetsparken 5, DK--2100 Copenhagen, Denmark.} \\
   \vspace{.4cm}
   {\large August 21, 2001} \\
   \vspace{.7cm}
   {\large\bf Abstract}
\end{center} 

\noindent
Motivated by the formalism of string bit models, or quantum matrix models, we
study a class of simple Hamiltonian models of quantum gravity type in two
space-time dimensions.  These string bit models are special cases of a more 
abstract class of models defined in terms of the $sl_2$ subalgebra of the 
Virasoro algebra.  They turn out to be solvable and their scaling limit 
coincides in special cases with known transfer matrix models of 
two-dimensional quantum gravity. 

\vspace{.5cm}

\begin{flushleft}
\end{flushleft}
\pagebreak

\section{Introduction}
\la{s1}

So far the most productive approach to two-dimensional quantum gravity has
been in terms of path integrals.  Specifically, the (Euclidean) path integral
formulation of Liouville gravity \cite{kpz} together with its regularized
versions involving random triangulations, or the equivalent matrix models (see,
e.g., Ref.\cite{adt}), have permitted detailed analysis of two-dimensional
quantum gravity coupled to matter fields with central charge $c \leq 1$, in
particular the loop correlation functions (or Hartle-Hawking wave functions) 
and, in some cases, the fractal characterstics of space-time.  Importantly, 
and as a rare instance in quantum field theory, the discrete models have 
provided analytic insights that presently seem out of reach of continuum 
methods, the most striking of which is perhaps the double scaling limit as a 
procedure to incorporate space-times of arbitrary topology.  (See 
Ref.\cite{adt} and the references therein.) 

One of the main motivations for studying quantum gravity in two dimensions is
based on the hope that it may serve as a testing ground for ideas and methods
extendable to higher dimensions. It is, indeed, straightforward to set up
discrete models of quantum gravity in arbitrary dimensions in terms of random
triangulations, but up to now very few analytic results have been obtained and
even very basic questions are left open. A number of numerical investigations
have, however, been carried out.  See, e.g., Ref.\cite{adt} and the references
therein.

The question arises, naturally, if there exist alternative formulations of the
regularized models, or closely related models, that are better tailored for
generalizations.  A second, and independent, purpose of such reformulations 
would be to make comparisons possible with continuum approaches other than the
path integral quantization, in particular canonical quantization \cite{ag}.  
In this paper we address this question in two dimensions and introduce a class
of (Euclidean) Hamiltonian models of regularized two-dimensional quantum 
gravity.  We do not claim to resolve the above mentioned questions, but we 
will show that the proposed models provide a Hamiltonian alternative to the 
discrete path integral (or transfer matrix) approach to a class of models 
introduced recently under the name {\it Lorentzian gravity} \cite{amlo, 
dgk99}, which on the other hand are closely related to the full randomly 
triangulated models mentioned at the beginning.  More precisely, we will show
that the continuum limits of the Lorentzian models can be obtained from our 
Hamiltonian models.  Indeed, we will define and solve a more abstract class 
of models and introduce a formalism opening up the way for even further 
generalizations.  Detailed analysis of these generalized models as well as 
generalizations to higher dimensions is, however, outside the scope of the 
present paper. 

Here is the basic idea of this work.  Recall that the primary goal of the 
Hamiltonian formulation of quantum gravity is to account for the 
``time''-evolution of a space-like  universe of fixed topology\footnote{Note 
that the models considered in this paper are all within the Euclidean 
framework.}.  Restricting first to connected and compact space-like universes 
implies that the two-dimensional space-time has either the topology of a 
strip, corresponding to equal-time slices that are open line segments, or a 
cylinder with circles as equal-time slices.  Since the only 
reparametrization-invariant quantity defined by a metric on a one-dimensional 
(connected) manifold is its volume, a natural way to discretise the spacial 
metric degree of freedom is to introduce a distance cutoff $a>0$ and consider 
the equal-time slices to be polygonal lines or loops, respectively, with 
volume $n \cdot a$, where the integer $n$ is the number of links in the
slice.  Keeping $a$ fixed, we associate with each such space-like universe of 
volume $n \cdot a$ a pure quantum state $| n \rangle$, and these are assumed to
form an orthogonal basis for the Hilbert space ${\cal T}$ of states.  The 
Hamiltonian acting on ${\cal T}$ is chosen in such a way that the action
couples adjacent links only.

It turns out that models of this type may conveniently be generated by a 
special variant of the string-bit formalism, whose basic ingredients are 
annihilation and creation operators that can be interpreted as annihilating or
creating links in the equal-time slices.  (We will explain the necessary 
details in Sections~\ref{s2} and \ref{s3}.)  String bit models were originally
developed as a means of regularizing string theory \cite{string, klsu}.  They 
provide suitable frameworks for quantum chromodynamics \cite{thorn} and 
quantum spin chain models \cite{9711052}, too.  Our variant may be envisaged
as generalized quantum spin chain models in which the numbers of ``spins'', 
i.e. links, are variable.  In this sense, the relationship between Lorentzian
gravity models and those string bit models equivalent to them is analogous to 
that between the six-vertex and the XXZ model \cite{baxter}; a better 
understanding of one class of models will spur the study of the other.

The paper is organised as follows.  In Section 2, the simplest possible 
Hamiltonian model in the case when space-time is a strip will be solved in
the continuum limit.  This model turns out to coincide with the corresponding
Lorentzian gravity model considered in \cite{dgk99};  in Section~\ref{s3}, 
we will consider a Hamiltonian model for cylindrical space-time which 
is not spacially homogeneous (or cyclically symmetric).  The model will be 
shown to reproduce, in the continuum limit, the Lorentzian model with a marked
link in the initial space-like slice considered in \cite{amlo}; in Section 4,
we will consider the cyclically symmetric version of the previously mentioned 
model and show that this, as well as the model in Section 2, can be obtained as
special cases of a more general class of models expressed in terms of the
$sl_2$-generators of the Virasoro algebra in a certain class of highest weight
representations, the Hamiltonian being of the form
\beq
   H = L_0 + \l L_1 + \l L_{-1}. 
\la{1.1}
\eeq
In Section~5, we will solve this model.  In particular, both the 
two-loop amplitude of the continuum Liouville gravity model in 
Ref.\cite{nakayama} and the so-called $p$-seamed correlation functions of 
Ref.\cite{dgk99} will be obtained as special cases; finally, we will discuss
briefly further generalisations and future developments in Section~\ref{s7}.

\section{Space-time with Boundaries}
\la{s2}

In this section, we are going to consider the quantum mechanics of a space-time
with the topology of a strip whose Hamiltonian is given by 
\beq
   H_{\frac{1}{2}} = {\rm Tr} \le\lb a^{\da} a + 
   \frac{\l}{\sqrt{N}} (a^{\da})^2 a + \frac{\l}{\sqrt{N}} a^{\da} a^2 \ri\rb
   - \frac{1}{4} \bar{q}^{\da} \bar{q} - \frac{1}{4} (q^{\da})^t q^t  
\la{2.1}
\eeq
for $N=\infty$.  Before explicating the various terms in Eq.(\ref{2.1}), let 
us briefly review what string bit models are.  We will largely follow 
Refs.\cite{9712090} and \cite{9806002}, {\em with a few modifications in 
definitions and notations.}

Consider an $N \times N$ matrix of creation operators.  Its matrix entry is 
written as $a^{\da\m_2}_{\m_1}$, where $\m_1$ and $\m_2$ are row and column 
indices, respectively, and can take any integer values between 1 and $N$ 
inclusive.  The Hermitian conjugate of this matrix is an $N \times N$ matrix 
of annihilation operators whose matrix entries are written in the form 
$a^{\m_1}_{\m_2}$.  The creation and annihilation operators satisfy the 
canonical commutation relations
\[ \le\lb a^{\m_1}_{\m_2}, a^{\da\m_3}_{\m_4} \ri\rb = \d^{\m_1}_{\m_4}
   \d^{\m_3}_{\m_2}. \]
All other commutators involving these operators vanish.

In addition, consider a $1 \times N$ row vector and an $N \times 1$ column 
vector of creation operators.  Their components are written as 
$\bar{q}^{\da\m}$ and $q^{\da}_{\m}$, respectively.   Their Hermitian 
conjugates are an $N \times 1$ column vector and a $1 \times N$ row vector of 
annihilation operators.  Their components take the form $\bar{q}_{\m}$ and 
$q^{\m}$, respectively.  These operators satisfy the canonical commutation 
relations
\[ \le\lb \bar{q}^{\da\m_1}, \bar{q}_{\m_2} \ri\rb = \d^{\m_1}_{\m_2}\quad
\mbox{and}\quad 
 \le\lb q^{\da}_{\m_1}, q^{\m_2} \ri\rb = \d^{\m_2}_{\m_1}. \]
All other commutators involving them, including those involving  
$a^{\da\m_2}_{\m_1}$ or $a^{\m_1}_{\m_2}$ as well, vanish.

Let $| \Om \rangle$ be a vacuum state annihilated by all annihilation 
operators, and define ${\cal T}_{\frac{1}{2}}$ as the Hilbert space spanned by
all states of the form
\bea
   | n \rangle_{\frac{1}{2}} & = & \frac{1}{N^{(n+1)/2}} \bar{q}^{\da\m_1} 
   a^{\da\m_2}_{\m_1} a^{\da\m_3}_{\m_2} \cd a^{\da\m_{n+1}}_{\m_n} 
   q^{\da}_{\m_{n+1}} | \Om \rangle \\
   & = & \frac{1}{N^{(n+1)/2}} \bar{q}^{\da} (a^{\da})^n q^{\da} | \Omega
   \rangle,
\eea
where $n$ is a positive integer and the summation convention is adopted for
all row and column indices.  The choice of notation $| \;\rangle_\frac{1}{2}$ 
for vectors in ${\cal T}_{\frac{1}{2}}$ will be explained in Section 4. The 
inner product $\langle \cdot \mid \cdot \rangle_{\frac{1}{2}}$ on 
${\cal T}_{\frac{1}{2}}$ is fixed uniquely by the commutation relations and 
$\langle \Om | \Om \rangle =1$.  Since \cite{9712090}  
\[ \lim_{N \rar \ift} \langle m | n \rangle_{\frac{1}{2}} = \d_{mn}, \]
the states $| 1 \rangle_{\frac{1}{2}}$, $| 2 \rangle_{\frac{1}{2}}$, \ld, and 
so on form an orthonormal basis of ${\cal T}_{\frac{1}{2}}$ in the large-$N$ 
limit, which is the limit we are considering.  We think of 
$| n \rangle_{\frac{1}{2}}$ as the quantum state of a universe made up of $n$ 
interior links and two boundary links. 

There are various kinds of natural operators acting on 
${\cal T}_{\frac{1}{2}}$.  We define
\beq
   \s^k_\ell & = & \frac{1}{N^{(k+\ell-2)/2}} a^{\da\m_2}_{\m_1} 
   a^{\da\m_3}_{\m_2} \cd a^{\da\n_\ell}_{\mu_k} a^{\n_{\ell-1}}_{\n_l} 
   a^{\n_{l-2}}_{\n_{\ell-1}} \cd a^{\m_1}_{\n_1} \nn \\
   & = & \frac{1}{N^{(k+\ell-2)/2}} {\rm Tr} \le\lb (a^{\da})^k a^\ell \ri\rb,
\la{2.2} 
\eeq
where $k$ and $\ell$ are any positive integers, and Tr denotes the trace in 
index space (and not on ${\cal T}_{\frac{1}{2}}$).  Moreover, we let
\beq
   l^0_0 = \bar{q}^{\da} \bar{q} 
\la{2.3}
\eeq
and 
\beq 
   r^0_0 = q^{\da}_{\m} q^{\m} = (q^{\da})^t q^t, 
\la{2.4}
\eeq
where the superscript $t$ denotes transposition.  In the large-$N$ limit 
\cite{thorn, 9712090},
\beq
   \s^k_\ell | n \rangle_{\frac{1}{2}} & = & \le\{ \ba{ll} 0 & \mbox{if} \; 
   \ell > n; 
   \\ (n-\ell+1) | n-\ell+k \rangle_{\frac{1}{2}} + {\cal O}(\frac{1}{N}) & 
   \mbox{if} \; \ell \leq n; \ea \ri. \nn \\
   l^0_0 | n \rangle_{\frac{1}{2}} & = &  | n \rangle_{\frac{1}{2}} \nn \\ 
   r^0_0  | n \rangle_{\frac{1}{2}} & = &  | n \rangle_{\frac{1}{2}},
\la{2.5}
\eeq
where ${\cal O}(1/N)$ consists of terms whose norms and whose inner products 
with any $| n \rangle_{\frac{1}{2}}$ are of the order of at most $1/N$.  These 
terms are thus negligible in the large-$N$ limit.  We can see from 
Eq.(\ref{2.5}) that $\s^k_\ell$ replaces any $\ell$ adjacent interior links 
with $k$ adjacent interior links and annihilates $| n \rangle_{\frac{1}{2}}$ 
if $\ell > n$.  $l^0_0$ and $r^0_0$ annihilate the left and right boundary 
links, respectively, and then create them back.  Hence, both $l^0_0$ and 
$r^0_0$ effectively act as the identity operator on 
${\cal T_{\frac{1}{2}}}$, but we will nevertheless display them explicitly 
below.

Using Eqs.(\ref{2.2}), (\ref{2.3}), and (\ref{2.4}), we can rewrite the
Hamiltonian $H_{\frac{1}{2}}$ in Eq.(\ref{2.1}) as
\beq 
   H_{\frac{1}{2}} = \s^1_1 + \l \s^2_1 + \l \s^1_2 - \frac{1}{4} l^0_0 
   - \frac{1}{4} r^0_0, 
\la{2.6}
\eeq
where $\l$ is a real constant.  In this formula, $\s^1_1$ is the interior
spatial volume energy term.  Each interior link carries one unit of energy,
and the volume  energy of a state is proportional to the number of links.
$\s^2_1$ splits any interior link into two.  Since $| n \rangle_{\frac{1}{2}}$ 
represents $n$ interior links, $\s^2_1$ maps $| n \rangle_{\frac{1}{2}}$ to 
$n| n+1 \rangle_{\frac{1}{2}}$.  On the other hand, $\s^1_2$ combines any two 
adjacent interior links into one.  Since there are only $n-1$ pairs of 
adjacent interior links in $| n \rangle_{\frac{1}{2}}$, $\s^1_2$ maps 
$| n \rangle_{\frac{1}{2}}$ to $(n-1)| n \rangle_{\frac{1}{2}}$ . Finally, the 
last two terms represent boundary volume energy, but notice that the two 
boundary links contribute negative energy, in total minus one half the energy 
of an interior link.  We stress that this value of the  relative size of the 
volume energy contributions is crucial for the existence of a continuum limit 
as will be seen.  By now, it should have been obvious that $H_{\frac{1}{2}}$, 
featuring the physics of spatial homogeneity and locality, is among the 
simplest Hamiltonians one can conceive for a spacetime with boundaries.

We now proceed to evaluate the transition amplitudes 
\[ \ti{G}_{\frac{1}{2}} (k, l; t) = \langle l | e^{-t H_{\frac{1}{2}}} 
   | k \rangle_{\frac{1}{2}} \]
in the continuum limit. It is convenient to work with its generating function
\[ G_{\frac{1}{2}} (x, y; t) = \sum_{k,l=1}^{\ift} x^k y^l 
   \ti{G}_{\frac{1}{2}} (k, l; t), \]
where $x, y$ are complex variables.  Introducing
\[ \ti{\T} (k, l; n) = \langle l | H_{\frac{1}{2}}^n | k 
   \rangle_{\frac{1}{2}} \]
and its generating function 
\[ \T (x, y; n) = \sum_{k,l=1}^{\ift} x^k y^l \ti{\T} (k, l; n), \]
for non-negative integers $n$, we have
\beq
    G_{\frac{1}{2}} (x, y; t) & = & \sum_{n=0}^{\ift} \frac{(-t)^n}{n!} 
    \T (x, y; n).
\la{2.14}
\eeq
Note, firstly, from
\[ \ti{\T} (k, l; 1) = \langle l | H_{\frac{1}{2}} | k 
   \rangle_{\frac{1}{2}} = \le( k - \frac{1}{2} \ri) \d_{lk} + \l k 
   \d_{l, k+1} + \l \le( k - 1 \ri) \d_{l, k-1} \]
that 
\beq
   \T (x, y; 1) = \frac{xy (1 + \l x + \l y)}{(1 - xy)^2} - 
                  \frac{xy}{2 (1 - xy)}. 
\la{2.8}
\eeq
Secondly, since  $| 1 \rangle_{\frac{1}{2}}$, $| 2 
\rangle_{\frac{1}{2}}$, \ld, and so forth form an orthonormal basis of
${\cal T}_{\frac{1}{2}}$, we have \cite{amlo}
\beq
   \T (x, y; n) = \oint \frac{dz}{2 \pi i z} \T (x, \frac{1}{z}; 1)
   \T (z, y; n-1), 
\la{2.9}
\eeq
where both $\T(x, z^{-1}; 1)$ and $\T (z, y; n-1)$ are treated as complex 
functions of $z$, and the contour encirles all singularities of 
$\T (x, z^{-1}; 1)$ but none of $\T (z, y; n-1)$.  Using Eq.(\ref{2.8}), we
find that Eq.(\ref{2.9}) leads to
\[ \T (x, y; n) = \le\lb - \le( \frac{1}{2} + \frac{\l}{x} \ri) + 
   \le( \l + x + \l x^2 \ri) \frac{\partial}{\partial x} \ri\rb
   \T (x, y;n-1). \]
By Eq.(\ref{2.14}), this yields 
\beq
   \frac{\partial}{\partial t} G_{\frac{1}{2}} (x, y; t) + 
   \le\lb - \le( \frac{1}{2} + \frac{\l}{x} \ri) + 
   \le( \l + x + \l x^2 \ri) \frac{\partial}{\partial x} \ri\rb 
   G_{\frac{1}{2}} (x, y; t) = 0. 
\la{2.15}
\eeq
Together with the initial condition
\beq
   G_{\frac{1}{2}} (x, y; 0) =  \T (x, y; 0) = \frac{xy}{1-xy}
\la{2.16}
\eeq
this first order partial differential equation determines $G_{\frac{1}{2}} 
(x, y; 0)$ uniquely.

We are presently only interested in evaluating the continuum limit of
$G_{\frac{1}{2}}$. Singularities appear when the coefficients in the square 
brackets in Eq.(\ref{2.15}) vanish, i.e., for $x = y = \pm 1$ and $\l = \mp
\frac{1}{2}$.  These are identical to the critical values found in 
Refs.\cite{amlo} and \cite{dgk99}, and we can apply the same scaling 
procedure.  Hence we set
\beq
   t = \frac{2T}{a}, \; x = e^{-Xa}, \; y = e^{-Ya}, \; \mbox{and} \; 
   \l = - \frac{1}{2} e^{- \frac{\La}{2} a^2},
\la{2.17}
\eeq
where $a$ is the distance cutoff and $T$, $X$, $Y$, and $\La$ are finite
renormalized values of $t$, $x$, $y$, and $\l$, respectively, and define the
continuum continuum limit of $G_{\frac{1}{2}}$, for which we use the same 
notation, by 
\beq
   G_{\frac{1}{2}} (X, Y; T) = \lim_{a \rar 0} a G_{\frac{1}{2}} (x, y; t)\;. 
\la{2.18}
\eeq
Substituting Eq.(\ref{2.17}) into Eqs.(\ref{2.15}) and (\ref{2.16}) yields the
limiting equation 
\[ \frac{\partial}{\partial T} G_{\frac{1}{2}} (X, Y; T) + 
   \le( X^2 - \La \ri) \frac{\partial}{\partial X} G_{\frac{1}{2}} (X, Y; T) +
   X G_{\frac{1}{2}} (X, Y; T) = 0  \]
with the initial condition
\[ G_{\frac{1}{2}} (X, Y; 0) = \frac{1}{X+Y}. \]

These are identical to the equations found for a model of Lorentzian gravity
with boundaries in Ref.\cite{dgk00}. By inverse Laplace transformation of the 
solution with respect to $X$ and $Y$, one finds (see Ref.\cite{amlo}) the 
continuum limit of the transition amplitude expressed in terms of the physical
lengths $L = k \cdot a, L' = l \cdot a$ of the two boundary components to be
\beq
   \ti{G}_{\frac{1}{2}} (L, L'; T) = \frac{\sqrt\Lambda}{\sinh(T\sqrt\Lambda)}
   e^{-\sqrt\Lambda(L+L')\coth(T\sqrt\Lambda)}I_0\left(
   \frac{2\sqrt{\Lambda L L'}}{\sinh(T\sqrt\Lambda)}\right)\;,
\la{G1}
\eeq
where $I_0$ is the zeroth modified Bessel function.  Consequently, our string 
bit model and this particular model of Lorentzian gravity belong to the same 
universality class.  We will come back to this model again in Section~\ref{s4}.

\section{Closed, Non-homogeneous Space-time}
\la{s3}

In this section we discuss an example of a Hamiltonian model of cylindrical
space-time. Conceptually, it is not the simplest such model, which we defer to
the next section. However, it has the virtue of being solvable by the 
generating function technique of the preceding section, which is our main
motivation for considering it here. The model is spatially non-homogeneous in 
the sense that the equal time slices have one marked link, that is a link 
created by a matrix of creation operators different from those creating the 
rest. A model of Lorentzian gravity with a marked initial loop has been 
studied in Ref.\cite{amlo}, and we will find that its continuum limit is 
reproduced by our model.
 
The Hamiltonian we consider is given by
\beq
   H_c & = & {\rm Tr} ( a^{\da} a + b^{\da} b ) + 
   \frac{\l}{\sqrt{N}} {\rm Tr} ( a^{\da} a^{\da} a + 
   \frac{1}{2} a^{\da} b^{\da} b +
   \frac{1}{2} b^{\da} a^{\da} b \nn \\ 
   & & + a^{\da} a^2 + b^{\da} b a + b^{\da} a b )
\la{3.2}
\eeq
for $N=\infty$.  As mentioned, this quantum matrix model requires a second
matrix of creation operators besides $a^{\da}$.  The entries of this matrix 
are written as $b^{\da\m_1}_{\m_2}$, whose corresponding annihilation operator
is $b^{\m_2}_{\m_1}$.  They satisfy the same canonical commutation relations as
the $a$-operators and commute with these.

Let ${\cal T}_c$ be the Hilbert space spanned by all states of the form
\[ | n \rangle_c =  \frac{1}{N^{n/2}} {\rm Tr} \le\lb b^{\da} 
   (a^{\da})^{n-1} \ri\rb | \Omega \rangle, \]
where $n$ is an arbitrary positive integer.  These states form an orthonormal 
basis for ${\cal T}_c$ in the large-$N$ limit \cite{thorn, 9712090}:
\[ \lim_{N \rar \ift} \langle m | n \rangle_c = \d_{mn}. \]
We consider $| n \rangle_c$ as the state of a closed universe with $n$ 
links, one of which, created by $b^{\da}$, is marked.  (c.f. the string bit
interpretation in Refs.\cite{string} and \cite{klsu}.)

The operators acting on ${\cal T}_c$ which are relevant to us may be
written either as
\beq
   g^k_l = \frac{1}{N^{(k+l-2)/2}} {\rm Tr} \le\lb (a^{\da})^k a^l \ri\rb,
\la{3.3}
\eeq
where $k$ and $l$ are positive integers, or as
\beq
   g^{k_1,k_2}_{l_1,l_2} = \frac{1}{N^{(k_1 + k_2 + l_1 + l_2)/2}} 
   {\rm Tr} \le\lb (a^{\da})^{k_1} b^{\da} (a^{\da})^{k_2} 
   a^{l_2} b a^{l_1} \ri\rb,
\la{3.4}
\eeq
where $k_1$, $k_2$, $l_1$, and $l_2$ are non-negative integers.  ($\s^k_l$ and
$g^k_l$ are the restrictions of the same operator to the open and closed string
sectors, respectively.  They are elements of different Lie algebras 
\cite{9712090, 9806002}, a fact we will see and use in Section~\ref{s4}.)  In 
the large-$N$ limit \cite{thorn, 9712090}, 
\[ g^k_l | n \rangle_c = \le\{ \ba{ll} 0 & \mbox{if} \; l \geq n \;
   \mbox{or} \\ 
   \le( n - l \ri) | {n-l+k} \rangle_c + {\cal O}(\frac{1}{N}) & 
   \mbox{if} \; l < n, \ea \ri. \]
and
\[ g^{k_1,k_2}_{l_1,l_2} | n \rangle_c = \le\{ \ba{ll} 0 & \mbox{if} \; 
   l_1 + l_2 > n-1 \; \mbox{or} \\ 
   | n - l_1 - l_2 + k_1 + k_2 \rangle_c + {\cal O}(\frac{1}{N}) & 
   \mbox{if} \; l_1 + l_2 < n. \ea \ri. \]
Thus, $g^k_l$ replaces any adjacent $l$ unmarked links in $| n 
\rangle_c$ with $k$ unmarked links, and $g^{k_1,k_2}_{l_1,l_2}$ replaces 
adjacent $l_1 + l_2 + 1$ links, where the $(l_1 + 1)$-th link is marked, with 
$k_1 + k_2 + 1$ links, where the $(k_1 + 1)$-th link is marked.  Note that 
these operators always preserve the marked link.

Using Eqs.(\ref{3.3}) and (\ref{3.4}), we can paraphrase $H_c$ in 
Eq.(\ref{3.2}) as
\[ H_c = g^1_1 + g^{0,0}_{0,0} + \l \le( g^2_1 + \frac{1}{2} g^{1,0}_{0,0} + 
   \frac{1}{2} g^{0,1}_{0,0} + g^1_2 + g^{0,0}_{1,0} + g^{0,0}_{0,1} \ri), \]
where $\l$ is a real constant.  In this formula, $g^1_1 + g^{0,0}_{0,0}$ is 
the volume energy term. The terms $g^2_1 + 1/2 g^{1,0}_{0,0} + 1/2
g^{0,1}_{0,0}$ implement splitting of any unmarked link into two unmarked 
links or splitting the marked link into a marked and an unmarked link.  
Finally, the terms $g^1_2 + g^{0,0}_{1,0} + g^{0,0}_{0,1}$ combine a pair of 
juxtaposed unmarked links into one unmarked link or combine the marked link 
and a juxtaposed unmarked link into the marked link. Notice that the relative 
constants of the terms in $H_c$ are chosen such that its action treats the 
marked link in the same way as the unmarked ones. More specifically, 
\[ \le( g^1_1 + g^{0,0}_{0,0} \ri) |  n \rangle_c = n | n \rangle_c \;,\]
\[ \le( g^2_1 + \frac{1}{2} g^{1,0}_{0,0} + \frac{1}{2} g^{0,1}_{0,0} \ri)
   | n \rangle_c = n | n + 1 \rangle_c \;, \]
\[ \le( g^1_2 + g^{0,0}_{1,0} + g^{0,0}_{0,1} \ri) | n \rangle_c = 
   n | n - 1 \rangle_c. \]
Thus this form of $H_c$ appears to represent the most natural nearest 
neighboring interaction on ${\cal T}_c$, but notice that it is 
non-Hermitian.   

As already mentioned, it turns out that the model so defined can be solved
by the same method as that of the preceding section. Since the differences
between the calculations are only minor, we will skip the details. Using the
scaling conditions (\ref{2.17}) one finds that the the continuum limit
$G_c (X,Y;T)$, defined by the same procedure as in Section 2, fulfills
\[ \frac{\partial}{\partial T} G_c (X, Y; T) + 
   \le( X^2 - \La \ri) \frac{\partial}{\partial X} G_c (X, Y; T) +
   2X G_c (X, Y; T) = 0   \]
with the initial condition
\[ G_c (X, Y; 0) = \frac{1}{X+Y}. \]
These equations are identical to those found for a Lorentzian gravity model in
Ref.\cite{amlo} with one marked link in the entrance loop.  Consequently, the 
two continuum limits coincide, as claimed. For later reference we note that 
the solution in terms of the length variables is \cite{amlo} 
\beq
   \ti{G}_c (L, L'; T) = \sqrt{\frac{L}{L'}} 
   \frac{\sqrt{\La}}{\sinh (\sqrt{\La} T)}
   e^{- \sqrt{\La} (L + L') \coth (\sqrt{\La} T)}
   I_1 \le( \frac{2 \sqrt{\La L L'}} {\sinh (\sqrt{\La} T)} \ri).
\la{3.8}
\eeq

\section{Closed, Homogeneous Space-time and Tensor Product models}
\la{s4}

Marking a  link in a boundary loop is a convenient technical device in
triangulated models and the relation to the same model with no marking is
simple, since the marking only gives rise to a factor equal to the length of
the marked loop in the counting of triangulations.  The relation is of a 
different kind for Hamiltonian models but, as we will immediately see, still 
quite straightforward. 

In spatially homogeneous models, all links in an equal-time slice have
identical status, meaning that only one type of creation and annihilation
operators is involved. The simplest nearest neighboring Hamiltonian is then
given by
\beq
   H_1 = {\rm Tr} \le\lb a^{\da} a + 
   \frac{\l}{\sqrt{N}} (a^{\da})^2 a + \frac{\l}{\sqrt{N}} a^{\da} a^2 \ri\rb
\la{4.1}
\eeq
with $N=\infty$ and $\l$ a real parameter.  Comparing Eqs.(\ref{3.2}) and 
(\ref{4.1}), we see that removing the marked link restores not only cyclic 
symmetry but also Hermiticity.

The Hilbert space ${\cal T}_1$ on which $H_1$ acts is spanned by all states of
the form
\[ | n \rangle_1 = \frac{1}{N^{n/2}} {\rm Tr} (a^{\da})^n | \Om 
   \rangle, \]
where $n$ is a positive integer.  In the large-$N$ limit \cite{thorn, 9712090},
\[ \lim_{N \rar \ift} \langle m | n \rangle_1 = n\delta_{mn}. \]
Hence $| 1 \rangle_1$, $| 2 \rangle_1$, \ldots, and so on form an orthogonal 
basis for ${\cal T}_1$.  We think of $| n \rangle_1$ as the quantum state of a 
closed one-dimensional universe topologically equivalent to a regular polygon 
with $n$ sides (c.f. the string bit interpretation in Refs.\cite{string} and 
\cite{klsu}).  In terms of the operators introduced in Eq.(\ref{3.3}), the 
Hamiltonian $H_1$ can be rewritten as
\[ H_1 = g^1_1 + \l g^2_1 + \l g^1_2, \]
In the large-$N$ limit \cite{thorn, 9712090},
\bea
   g^1_1 | n \rangle_1 & = & n | n \rangle_1 \; \mbox{if} \; n \geq 0, \\
   g^2_1 | n \rangle_1 & = & n | n + 1 \rangle_1 \; \mbox{if} \; n \geq 0, \\
   g^1_2 | 1 \rangle_1 & = & 0, \; \mbox{and} \\
   g^1_2 | n \rangle_1 & = & n | n - 1 \rangle_1 \mbox{if} \; n > 0. 
\eea
Here $g^1_1$ is again a volume energy term, $g^2_1$ splits any link into two 
and $g^1_2$ combines any pair of adjacent links into one. 

It turns out to be tricky, if not impossible, to apply the previously used 
generating function technique to work out the transition amplitude of this 
Hamiltonian in the continuum limit.  Instead, we will derive it by 
diagonalising $H$.  Before doing so, however, we will digress for a moment and
make some observations about the underlying Lie algebras of string bit models.
This will lead to the introduction of a more general and abstract class of 
Hamiltonian models including the one just defined as well as the model in 
Section 2 and, in the continuum limit, the tensor product type models of 
Ref.\cite{dgk99} as special examples.

It was shown in Ref.\cite{9806002} that if we take ${\cal T}_1$ as the 
defining representation of the Lie algebra $\hat{C}_1$ generated by all 
$g^k_l$'s, then under the identifications\footnote{There is an important
difference between the way the Virasoro generators arose in 
Refs.\cite{9806002} and \cite{9712090} and the way they arise here.  In those 
two articles, every Virasoro generator $L_n$ was identified as a {\em coset} 
of certain elements of the Lie algebra $\hat{C}_1$ or $\hat{\S}_1$, a 
subalgebra of $\hat{G}_{1,1}$.  These cosets satisfied the Witt algebra, i.e, 
the classical Virasoro algebra.  Therefore, the Witt algebra was a {\em 
quotient algebra} of $\hat{C}_1$ or $\hat{\S}_1$.  Here, on the other hand, 
$L_{-1}$, $L_0$ and $L_1$ (and nothing else) are identified with {\em specific
elements} of $\hat{C}_1$ or a variant of $\hat{\S}_1$.  They turn out to form 
the $sl_2$ subalgebra of the Virasoro algebra. Therefore, $sl_2$ is a 
{\em subalgebra} of $\hat{C}_1$ or the variant of $\hat{\S}_1$.}
\[ g^1_1 \lrar L_0, \; g^1_2 \lrar L_1, \; \mbox{and} \; g^2_1 \lrar L_{-1}, \]
they satisfy the Lie brackets 
\beq
   \le\lb L_0, L_1 \ri\rb & = & - L_1, \nn \\
   \le\lb L_0, L_{-1} \ri\rb & = & L_{-1}, \; \mbox{and} \nn \\
   \le\lb L_1, L_{-1} \ri\rb & = & 2 L_0,
\la{6.2}
\eeq
which the reader may verify directly and easily.  This Lie algebra is nothing 
but the $sl_2$ subalgebra of the Virasoro algebra.  Furthermore, since
\[ g^1_2 | 1 \rangle_1 = 0, \; g^1_1 | 1 \rangle_1 = 
  | 1 \rangle_1 \; \mbox{and} \; \langle 1 | 1 \rangle_1 = 1, \]
$| 1 \rangle_1$ plays the role of the highest weight vector in the 
defining representation, and the highest weight $h = 1$.  It should thus be 
natural for us to consider the model in which the Hamiltonian 
\[ H = L_0 + \l L_{-1} + \l L_1 \]
is an element of $sl_2$ and acts on a certain highest weight represetation.  
Recall that for general $h>0$ the representation space ${\cal T}_h$ is the 
Hilbert space spanned by the vectors 
\beq 
   | n+1 \rangle_h = \frac{1}{n!} L_{-1}^n | 1 \rangle_h,
\la{6.4}
\eeq
where $| 1 \rangle_h$ is the highest weight vector and $n$ is any 
non-negative integer.  The actions of $L_{-1}$, $L_0$, and $L_1$ are given by
\beq
   L_0 | n+1 \rangle_h  & = & 
   (n+h) | n+1 \rangle_h \; \mbox{if} \; n \geq 0, \nn \\
   L_{-1} | n+1 \rangle_h & = & (n+1) | n+2 \rangle_h \; 
   \mbox{if} \; n \geq 0, \nn \\
   L_1 | 1 \rangle_h & = & 0, \; \mbox{and} \nn \\
   L_1 | n+1 \rangle_h & = & (n-1+2h) | n \rangle_h \; 
   \mbox{if} \; n > 0.
\la{6.6}
\eeq
The inner product on ${\cal T}_h$ is uniquely determined by the commutation
relations (\ref{6.2}), $\langle 1 | 1 \rangle_h = 1$, $L_0^{\da} = L_0$, and 
$L_1^{\da} = L_{-1}$. In particular, 
\beq
   \langle n+1 | n+1 \rangle_h = \frac{\G (n+2h)}{n! \G (2h)}.
\la{6.5}
\eeq
We note that the inner product is positive-definite if and only if $h>0$.

Next, we revisit the model in Section~\ref{s2} for a spacetime with 
boundaries.  Its Hamiltonian $H_{\frac{1}{2}}$ was given by Eq.(\ref{2.6}).  
The following observation is a slight modification of the results concerning
the Lie algebra $\hat{\S}_1$ generated by all $\s^k_l$'s in 
Ref.\cite{9712090}: make the identifications\footnote{{\em ibid}.}
\[ \s^1_1 - \frac{1}{2} l^0_0 \lrar L_0, \; \s^1_2 \lrar L_1, \; \mbox{and} \; 
   \s^2_1 \lrar L_{-1}. \]
Then the actions of $L_{-1}$, $L_0$, and $L_1$ on ${\cal T}_{\frac{1}{2}}$ 
satisfy the Lie brackets (\ref{6.2}).   We are thus again led to the 
Hamiltonian $H$ of the form (\ref{1.1}).  Since
\[ \s^1_2 | 1 \rangle_{\frac{1}{2}} = 0, \;
   \le( \s^1_1 - \frac{1}{2} l^0_0 \ri) | 1 \rangle_{\frac{1}{2}} = 
   \frac{1}{2} | 1 \rangle_{\frac{1}{2}} \; \mbox{and} \; 
   \langle 1 | 1 \rangle_{\frac{1}{2}} = 1, \]
$| 1 \rangle_{\frac{1}{2}}$ plays the role of a normalised highest weight 
vector.

Based on these observations, we will, in the following, consider Hamiltonians
of the form (\ref{1.1}) for arbitrary positive highest weights $h$. It turns
out to be possible to diagonalise $H$ for all such $h$ and to evaluate the
continuum limit of the transition amplitude. Before we perform this task in
the next section, a few remarks on the interpretation of these models are in
order. 

As is well known from the representation theory of $sl_2$, its $h = 1$ highest
weight representation is the symmetric tensor product of two copies of the 
$h = 1/2$ highest weight representation.  This fact has a clear interpretation
in terms of the gravity models as follows.  Taking two copies of the 
Hamiltonian $H_{\frac{1}{2}}$, we define the symmetrised tensor product state 
\[ | n \rangle' = \sum_{k=1}^n | k\rangle_{\frac{1}{2}} \otimes
   | n-k+1 \rangle_{\frac{1}{2}} \]
for $n\geq 1$ and regard it as representing the state of a closed polygon
obtained by gluing two polygonal lines of total length $n$ at both ends, where
each of the four boundary links of the two polygonal lines has a negative 
length of $-1/4$.  The action of $H_{\frac{1}{2}} \otimes 1+1 \otimes 
H_{\frac{1}{2}}$ on the space spanned by the vectors $| n\rangle$ will then be
easily seen to equal to that of $H$ for $h=1$ under the identification 
$| n \rangle' = | n \rangle$.  Similar remarks apply to the generators 
$L_{-1}$, $L_0$, and $L_1$.

Corresponding considerations of tensor products and gluing constructions for
Lorentzian gravity models were discussed in Ref.\cite{dgk99}, as were 
extensions to multiple tensor products. The latter lead to the so-called 
$p$-seamed transition amplitudes. As will also be seen from the explicit 
solution in the next section, these are reproduced by our algebraic model for 
integer values of $2h$.  

\section{Solution to $sl_2$ Gravity Model}
\la{s6}

In this section we will show how to obtain the continuum limit of the models 
with Hamiltonian given by Eq.(\ref{1.1}) for arbitrary $h>0$. We will first 
prove that $H$ is diagonalisable and determine the exact energy spectrum. Then
we will determine the asymptotic form of the eigenvectors close to 
criticality, which will enable us to extract the continuum limit.

As already remarked,  ${\cal T}_h$ is spanned by the vectors 
$| 1 \rangle_h$, $| 2 \rangle_h$, \ld, and so forth defined by 
Eq.(\ref{6.4}).  These vectors being orthogonal, it follows that states in 
${\cal T}_h$ are given by 
\beq
\sum_{n=1}^\ift a_n | n \rangle_h,
\la{ort}
\eeq
where
\beq
\sum_{n=1}^\ift n^{2h-1} |a_n|^2 < \infty
\la{norm}
\eeq
by Eq.(\ref{6.5}).  Clearly, $H$ is an unbounded operator defined, e.g., on 
vectors for which the sequence $a_n$ is rapidly decreasing.

\subsection{Diagonalisation of $H$}

We will apply a refined version of the Frobenius method used to solve a very 
similar quantum matrix model in Ref.\cite{9712090}.  Assume, for $E \geq 0$, 
that  
\beq
   \ph = \sum_{n=1}^{\ift} a_n | n \rangle_h 
\la{4.4.1}
\eeq
is an eigenstate of $H$.  Using Eqs.(\ref{1.1}) and (\ref{6.6}), we may 
write the eigenvalue equation 
\[ H \ph = E \ph \]
as
\[ \l (n+2h) a_{n+2} + (n+h-E) a_{n+1} + \l n a_n = 0 \]
for all non-negative values of $n$.  (The value of the new unknown $a_0$ is
immaterial because its coefficient is 0.)  Asymptotically for large $n$, the
equation reduces to
\[ \l a_{n+2} + a_{n+1} + \l a_n \simeq 0, \]
whose solutions are of the form
\[ a_n \simeq \alpha p^n + \beta p^{-n}, \]
where
\beq 
   p = \frac{-1 + \sqrt{1 - 4 \l^2}}{2 \l} 
\la{4.5.0.1}
\eeq
in the continuum limit.  Since $| p| < 1$ for $| \l | < 1/2$, it follows that 
$a_n$ must asymptotically behave as $p^n$ in order that $\ph$ be 
normalisable in accordance with Eq.(\ref{norm}).  Hence, we set
\[ a_n = b_n p^n, \]
resulting in the equation
\[ \l (n+2h) b_{n+2} p^2 + (n+h-E) b_{n+1} p + \l n b_n = 0 \]
for all non-negative values of $n$.

In terms of the increments
\[ \D b_n = b_{n+1} - b_n \quad\mbox{and}\quad
   \D^2 b_n = \D b_{n+1} - \D b_n, \]
this equation may be rewritten as
\bea
   & \l (n + 2h) p \D^2 b_n + 
   \le\lb (2 \l p + 1) n + (4h \l p + h - E) \ri\rb \D b_n & \\
   & + (2h \l p + h - E) b_n = 0. &
\eea
Introduce the ansatz \cite{spiegel}
\[ b_n = \sum_{r = 0}^{\ift} c_r n^{(r)}, \]
where the real numbers $c_r$ depend not on $n$ but on $r$ only, and the
factorial polynomial  $n^{(r)}$ is defined by 
\[ n^{(r)} = \le\{ \ba{ll} 
   n (n-1) \cd (n-r+1) & \mbox{if} \; r > 0; \; \mbox{and} \\
   1 & \mbox{if} \; r = 0\;. \ea \ri. \]
From
$$
   \D b_n = \sum_{r = 0}^{\ift} r c_r n^{(r-1)} 
\quad\mbox{and}\quad
   \D^2 b_n = \sum_{r = 0}^{\ift} r (r-1) c_r n^{(r-2)},
$$
we then obtain the equation
\beq
   & \l p (r+2) (r+1) (r+2h) c_{r+2} +
   \le\lb (3r+4h) \l p + (r+h-E) \ri\rb (r+1) c_{r+1} & \nn \\
   & + \le\lb (1 + 2 \l p) (r + h) - E \ri\rb c_r = 0 & 
\la{4.11}
\eeq
as a sufficient condition for the coefficients $c_r$ to fulfill
for all values of $r$. 

For a non-negative integer $R$, we now set
\beq
   E  = E_R = (1 + 2 \l p) (R + h) 
\la{4.12}
\eeq
and see that we obtain a unique solution for $c_r$, up to constant multiples,
for which $c_r=0$ for $r>R$. The corresponding eigenstate $\ph_R$ is thus of
the form
\[ \ph_R=\sum_{n=1}^\infty C_R(n)p^n| n\rangle_h, \]
where $C_R(n)$ is a polynomial of degree $R$ in $n$.

Since $H$ is Hermitian, the found eigenstates form an orthogonal set. Moreover,
it is a complete set, which can be seen as follows. Take any vector in
${\cal T}_h$ given by (\ref{ort}) and assume it is orthogonal to all
eigenvectors $\ph_R$. Since the $C_R(n)$'s span all polynomials, this means
that 
$$
\sum_{n=1}^\infty a_n n^S p^n  \frac{\G (n+2h)}{n! \G (2h)}=0\;,
$$
where $S$ is an arbitrary non-negative integer and we have used 
Eq.(\ref{6.5}).  Multiplying this equation by $z^S/S!$ and summing over $S$ 
give
$$
\sum_{n=1}^\infty a_n p^n \frac{\G (n+2h)}{n! \G (2h)}e^{zn}=0
$$
for $|z|<-\log p$. Obviously, the left hand side is an analytic function of $z$
in the half plane $\Re(z) < - \log p$ and hence vanishes there. Restricting
$z$ to the imaginary axis, we obtain a vanishing Fourier series, and
consequently its Fourier coefficients vanish. This proves that $a_n=0$ for all
$n\geq 1$ and the completeness of $\ph_R$ for all $R \geq 0$ follows.

Thus Eq.(\ref{4.12}) gives the whole energy spectrum.  Note that in the limit
(\ref{2.17}),
\[ E_R \simeq \sqrt{\La} a (h + R) \rar 0 \]
as $a \rar 0$, so the model is well behaved in this limit.

\subsection{Asymptotic behaviour of eigenstates}

In order to determine the continuum limit of the transition amplitude we will
need the asymptotic behaviour of the polynomials $C_R(n)$ under the scaling
conditions given in Eq.(\ref{2.17}).  Since $n$ scales as $a^{-1}$ we need to
exhibit the leading behaviour of the coefficients $c_r$ in $C_R(n)$ as $a \to
0$.  Make the ansatz that $c_{r+1}$ and $a c_r$ are of the same order in the
small-$a$ limit.  The recursion relation (\ref{4.11}) then gives
$$ c_{r+1} \simeq - \frac{R-r}{(r+1)(r+2h)} 2 \sqrt{\La} a c_r. $$
Iterating this equation yields 
$$
c_r\simeq c_0\frac{(-2 \sqrt{\La} a)^r \G(2h)}{\G(r+2h)} 
   \le( \ba{c} R \\ r \ea \ri)
$$
which, owing to the scaling of $n=L\cdot a^{-1}$, yields the asymptotic form 
\beq
   C_R(n)\simeq c_0\sum_{r=0}^R\frac{(-2 \sqrt{\La} a)^r \G(2h)}{\G(r+2h)} 
   \le( \ba{c} R \\ r \ea \ri) n^r,
\la{as1}
\eeq
where all summands are of order $1$.

The behavior of $c_0$ is fixed by requiring that $\ph_R$ be normalised. Using
the fact that $\ph_R$ is orthogonal to all vectors in ${\cal T}_h$ of the 
form (\ref{ort}) with $a_n = n^s p^n$, for $s$ = 0, 1, \ldots, and $R-1$, we 
get
\beq
  \langle \ph_R | \ph_R \rangle_h & = & \sum_{n=1}^\ift \sum_{m=1}^\ift
  \sum_{r=1}^R C_R(n) c_R m^R p^{n+m} \langle n | m \rangle_h \nn \\
  & = & \sum_{n=1}^\ift C_R(n) c_R n^R p^{2n} \frac{\G (n+2h)}{n! \G (2h)}.
\la{norm1}
\eeq
Substituting Eq.(\ref{as1}) into Eq.(\ref{norm1}) results in
\bea
  c_0^{-2} & \simeq & \sum_{r=0}^R \frac{(-2 \sqrt{\La} a)^{R+r} \G(2h)}
  {\G(R+2h) \G(r+2h)} \le( \ba{c} R \\ r \ea \ri) 
  \sum_{n=1}^{\ift} n^{R+r}  \frac{\G(n+2h)}{n!}p^{2n} \\ 
  & \simeq & \sum_{r=0}^R \frac{(-2 \sqrt{\La} a)^{R+r} \G(2h)}
  {\G(R+2h) \G(r+2h)} \le( \ba{c} R \\ r \ea \ri) 
  \sum_{n=1}^{\ift} \frac{\G(n+R+r+2h)}{n!} p^{2n}, 
\eea
where, in the last step, we have used that for $k$ real
$$
 \frac{\G(n+k)}{n!n^{k-1}}\to 1 \;\;\mbox{as}\;\;n\to\infty\;.
$$
By the binomial theorem and the relation
$$
p^2\simeq 1 - 2\sqrt{\La}a,
$$
we finally obtain
\beq
c_0^{-2} &\simeq& \sum_{r=0}^R \frac{(-2 \sqrt{\La} a)^{R+r} \G(2h)}
   {\G(R+2h) \G(r+2h)} \le( \ba{c} R \\ r \ea \ri) 
   \frac{\G(R+r+2h)}{(1-p^2)^{R+r+2h}}\nn \\
   & \simeq & \frac{1}{(2 \sqrt{\La} a)^{2h}} 
   \sum_{r=0}^R (-1)^{R+r} \frac{\G(R+r+2h) \G(2h)}{\G(R+2h) \G(r+2h)}
   \le( \ba{c} R \\ r \ea \ri) \nn \\
&=&  \frac{1}{(2 \sqrt{\La} a)^{2h}} \frac{R! \G(2h)}{\G(R+2h)}
\la{c0}\;,
\eeq  
where, in the last step, we have made use of the identity
$$
   \sum_{r=0}^R (-1)^{R+r} \frac{\G(R+r+2h)}{R! \G(r+2h)}
   \le( \ba{c} R \\ r \ea \ri) = 1\;,
$$
which is a special case of the Chu-Vandermonde identity.  (See, e.g., 
Ref.\cite{ls}.)

\subsection{The continuum limit}

We are now ready to compute the continuum limit of the transition amplitude.
The unnormalized transition amplitude is defined by
\beq
   \ti{G}_u (L, L'; T) = \lim_{a \rar 0} a^\alpha 
     \langle \frac{L'}{a} | e^{-t H} | \frac{L}{a} \rangle_h,
\la{4.18}
\eeq
where $t$ and $\l$ are given as in Eq.(\ref{2.17}), and the exponent $\a$ is 
to be determined such that the limit exists.  Recall from Eq.(\ref{6.5}) that 
the states $| \frac{L}{a} \rangle_h$ are not normalised.  On the other hand,
the more natural amplitude defined in terms of the normalised states may simply
be obtained from $\ti{G}_u$ by Eq.(\ref{6.5}); we will come back to it later.

Inserting the complete set of states $\{ \ph_R \}$, we have
\beq
   \ti{G}_u (L, L'; T) & = & \lim_{a \rar 0} a^\a \sum_{R=0}^\ift 
   e^{-tE_R} \langle \frac{L'}{a} | \ph_R \rangle_h \langle \ph_R | 
   \frac{L}{a}\rangle_h \nn \\
   & \simeq & \lim_{a \rar 0} a^\a \sum_{R=0}^\ift e^{-2 \sqrt{\La} (R+h)}
   \langle \frac{L'}{a}\mid\ph_R\rangle_h \langle\ph_R\mid \frac{L}{a}\rangle_h \;.
\la{cont1}
\eeq
Using Eqs.(\ref{as1}) and (\ref{c0}) as well as
$$ p^{\frac{L}{a}}\simeq e^{-\sqrt\La L}, $$
we have
\bea
   \lefteqn{\langle \ph_R | \frac{L}{a} \rangle_h =  
   C_R \le( \frac{L}{a} \ri) p^{\frac{L}{a}} \langle \frac{L}{a} 
   | \frac{L}{a} \rangle_h} \\
   & & \simeq a^{1-h} \le( 2 \sqrt{\La} \ri)^h 
   \sqrt{\frac{\G(R+2h)}{R!\G(2h)}} L^{2h - 1} 
   \sum_{r=0}^R \frac{(-2 \sqrt{\La})^r}{\G(r+2h)} 
   \le( \ba{c} R \\ r \ea \ri) L^r e^{- \sqrt{\La} L}. \nn
\eea
Inserting this expression into Eq.(\ref{cont1}) and choosing
\[ \a = 2h - 2, \]
we find that the continuum transition amplitude exists and takes the form
\beq
   \lefteqn{ \ti{G}_u (L, L'; T) = } \nn \\
   & & \le( 4 \La \ri)^h \le( L L' \ri)^{2h - 1} 
   \sum_{R=0}^{\ift} \frac{\G(R+2h)}{R!\G(2h)} \sum_{r=0}^R \sum_{s=0}^R
   \frac{(-2 \sqrt{\La})^{r+s} L^r L'^s}{\G(r+2h) \G(s+2h)} \nn \\
   & & \cdot \le( \ba{c} R \\ r \ea \ri) \le( \ba{c} R \\ s \ea \ri)
   e^{- \sqrt{\La} (L + L')} e^{- 2 (h+R) \sqrt{\La} T}.
\la{4.19}
\eeq

{\em A priori}, it is not obvious that this series is convergent for all 
positive values of $L$ ,$L'$, and $T$. One way to see this, and simultaneously
obtaining a more manageable expression for $\ti{G}_u$, is to apply an integral
representation of the reciprocal Gamma function (see, e.g., 
Ref.\cite{dettman}):
\beq
   \frac{1}{\G(x)} & = & \frac{1}{2 \pi i \b^{x-1}} 
   \int_{\ep - i \ift}^{\ep + i \ift} \frac{e^{\b (z+b)} dz}{(z+b)^x}.
\la{4.10}
\eeq
In this formula, $x$ and $\b$ are arbitrary positive numbers; $\ep$ is real 
and $b$ complex, and they satisfy $\Re(b) > \ep$; and the branch cut of 
$(z+b)^x$ lies on the negative real axis if $x$ is not an integer.  Apply 
Eq.(\ref{4.10}) to $1/\G(r+2h)$ with
\[ x = r+2h, \; \b = \sqrt{\La} L, \; b = 1, z = X, \]
and $\ep$ positive and infinitesimally small, and to $1/\G(s+2h)$ with
\[ x = s+2h, \; \b = \sqrt{\La} L', \; b = 1, \; z = Y, \]
and the same $\ep$ in Eq.(\ref{4.19}).  The binomial theorem can then be used 
to perform the summation over $r$ and $s$.  Apply once more the binomial 
theorem to the sum over $R$, and keep the branch cut of every $\lb \cd 
\rb^{2h}$ on the negative real axis.  With this choice of the branch cuts, 
$(z_1 z_2)^{2h} = z_1^{2h} z_2^{2h}$ if both $\Re(z_1)$ and $\Re(z_2)$ are 
positive.  Using this fact, we finally get
\bea
   \ti{G}_u (L, L'; T) & = &  4^h \La^{1-h}\frac{1}{2 \pi i} 
   \int_{\ep - i \ift}^{\ep + i \ift}  \frac{dY e^{\sqrt{\La} L' Y} 
   e^{-2h \sqrt{\La} T}}
   {\lb Y + 1 - (Y - 1) e^{-2 \sqrt{\La} T} \rb^{2h}} \nn \\
   & & \cdot \frac{1}{2 \pi i} \int_{\ep - i \ift}^{\ep + i \ift} 
   \frac{dX e^{\sqrt{\La} L X}}
   {\le\lb X + \frac{Y + 1 + (Y - 1) e^{-2 \sqrt{\La} T}}
   {Y + 1 - (Y - 1) e^{-2 \sqrt{\La} T}} \ri\rb^{2h}},
\eea
Integrals of this type have been carried out for integer values of $2h$ in 
\cite{dgk99}, but can be obtained for arbitrary values of $2h$. Indeed, apply 
Eq.(\ref{4.10}) to the integration with respect to $X$ and we immediately get
\bea
   \lefteqn{ \ti{G}_u (L, L'; T) = \frac{4^h e^{-2h \sqrt{\La} T}
   \sqrt{\La} L^{2h - 1}}{\G(2h) 2 \pi i}} \nn \\
   & & \cdot \int_{\ep - i \ift}^{\ep + i \ift} dY
   \frac{\exp \le\lb \sqrt{\La} L' Y - 
   \sqrt{\La} L \frac{Y (1 + e^{-2 \sqrt{\La} T})
   + (1 - e^{-2 \sqrt{\La} T})}{Y (1 - e^{-2 \sqrt{\La} T})
   + (1 + e^{-2 \sqrt{\La} T})} \ri\rb }
   { \lb Y (1 - e^{-2 \sqrt{\La} T}) + (1 + e^{-2 \sqrt{\La} T}) \rb^{2h}}.
\eea
Inserting the expansion
\bea
   & \exp \le\lb -\sqrt{\La} L\frac{Y (1 + e^{-2 \sqrt{\La} T})
   + (1 - e^{-2 \sqrt{\La} T})}{Y (1 - e^{-2 \sqrt{\La} T})
   + (1 + e^{-2 \sqrt{\La} T})} \ri\rb = \exp \le( - \sqrt{\La} L 
   \frac{1 + e^{-2 \sqrt{\La} T}}{1 - e^{-2 \sqrt{\La} T}} \ri) & \\
   & \cdot \sum_{n=0}^{\ift} \frac{1}{n!} 
   \le\lb \frac{4 \sqrt{\La} L e^{-2 \sqrt{\La} T}}{1 - e^{-2 \sqrt{\La} T}}
   \frac{1}{Y (1 - e^{-2 \sqrt{\La} T}) + (1 + e^{-2 \sqrt{\La} T})}
   \ri\rb^n, &
\eea
applying Eq.(\ref{4.10}) again to the remaining integration, and recalling that
\[ I_{2h-1}(z) = \sum_{n=0}^{\ift} \frac{z^{2n + 2h - 1}}{n! \G(2h + n)}. \]
is the $(2h-1)$-th modified Bessel function \cite{watson}, we finally obtain
$$ \ti{G}_u (L, L'; T) = \frac{(L L')^{h - \frac{1}{2}} \sqrt{\La}}
   {\G(2h) \sinh (\sqrt{\La} T)} e^{- \sqrt{\La} (L + L') \coth
   (\sqrt{\La} T)} 
   I_{2h-1} \le( \frac{2 \sqrt{\La L L'}} {\sinh (\sqrt{\La} T)} \ri). $$

Define the normalized transition function $\ti{G}(L, L'; T)$ by normalizing
$| \frac{L}{a} \rangle_h$ in Eq.(\ref{4.18}).  It follows from Eq.(\ref{6.5}) 
that we have to choose $\a = 1$ and that $\ti{G}(L, L'; T)$ deviates from 
${\ti G}_u (L, L'; T)$ by a factor of $(LL')^{h-\frac{1}{2}}/\Gamma(2h)$.
Consequently,
\beq
   \ti{G} (L, L'; T) = \frac{\sqrt{\La}}{\sinh (\sqrt{\La} T)} 
   e^{- \sqrt{\La} (L + L') \coth(\sqrt{\La} T)} 
   I_{2h-1} \le( \frac{2 \sqrt{\La L L'}} {\sinh (\sqrt{\La} T)} \ri).
\la{4.21} 
\eeq  

We note that if we put $h = 1/2$ in Eq.(\ref{4.21}), we will obtain 
Eq.(\ref{G1}); for $h = 1$ we obtain Eq.(\ref{3.8}) except for the factor 
$\sqrt{L/L'}$, which originates from the non-Hermiticity of $H_c$; if $h$ is
an integer, Eq.(\ref{4.21}) will coincide with the propagator calculated in 
the proper-time gauge of two-dimensional quantum gravity in Ref.\cite{nakayama}
with the winding number $h-1$; finally, if $2h$ is an integer, Eq.(\ref{4.21})
will be exactly the $(2h)$-seamed correlation function in Ref.\cite{dgk99}.

\section{Discussion}
\la{s7}

We have, in this paper, investigated different Hamiltonian models of
two-dimensional quantum gravity.  Clearly, one open problem is the proper
physical interpretation for the $sl_2$ gravity model with a non-integer value
of $2h$.  Perhaps this describe the interaction of gravity with matter, an
important future problem on its own right.  One could, for instance, study 
one-dimensional quantum spin systems whose Hamiltonians couple spin 
configurations of different sizes.  Diagonalisation of such Hamiltonians seems
to pose interesting new problems.  Another possibility is to extend the class 
of Hamiltonians defined in this paper by exploiting the representation theory 
of the full Virasoro algebra instead of the $sl_2$ subalgebra; such models 
might describe the coupling between matter and gravity.

Finally, extension to higher dimensional Hamiltonian models of quantum gravity
is an ultimate goal.  It is rather straightforward to produce candidate models
of nearest neighbouring type, but extracting non-trivial information from such
models seems a non-trivial task. We refer to Ref.\cite{amjulo} for a recent 
work on higher dimensional Lorentzian gravity. 

\vskip 1pc
\noindent \Large{\bf \hskip .2pc Acknowledgment}
\vskip 1pc
\noindent 

\normalsize
We thank J. Ambj{\o}rn and H. P. Jakobsen for discussions.

\end{document}